\documentclass[11pt]{amsart}
 \usepackage{amsopn}
 \usepackage{amsmath,amsthm,amssymb}

 \newcommand{\nc}{\newcommand}
 
  \nc{\aff}{\mathfrak{aff} } \nc{\bb}{\mathfrak{b} }
 \nc{\cc}{\mathfrak{c} }  \nc{\dd}{\mathfrak{d} } 
 \nc{\ggo}{\mathfrak{g} }
 \nc{\hh}{\mathfrak{h} }  \nc{\ii}{\mathfrak{i} }
 \nc{\jj}{\mathfrak{j} }  \nc{\kk}{\mathfrak{k} }
\nc{\mm}{\mathfrak{m} }   \nc{\nn}{\mathfrak{n} }
\nc{\pp}{\mathfrak{p} }  \nc{\rr}{\mathfrak{r} } \nc{\sg}{\mathfrak{s} }
 \nc{\sog}{\mathfrak{so} }  \nc{\spg}{\mathfrak{sp} }
 \nc{\sug}{\mathfrak{su} }  \nc{\slg}{\mathfrak{sl} }
 \nc{\tg}{\mathfrak{t} }  \nc{\uu}{\mathfrak{u} }
 \nc{\vv}{\mathfrak{v} } \nc{\ww}{\mathfrak{w} }
 \nc{\zz}{\mathfrak{z} }  
 
 \nc{\ggob}{\overline{\mathfrak{g}}} 
 
\nc{\glg}{\mathfrak{gl} }
  
\nc{\pca}{\mathcal{P}} \nc{\nca}{\mathcal{N}}
 
 \nc{\vp}{\varphi} \nc{\ddt}{\frac{{\rm d}}{{\rm d}t}}
 \nc{\la}{\langle} \nc{\ra}{\rangle}
 
 \nc{\SO}{{\sf SO}} \nc{\Spe}{{\sf Sp}} \nc{\Sl}{{\sf Sl}}
 \nc{\SU}{{\sf SU}} \nc{\Or}{{\sf O}} \nc{\U}{{\sf U}}
 \nc{\Gl}{{\sf Gl}} \nc{\Se}{{\sf S}} \nc{\Cl}{{\sf Cl}}
 \nc{\Spin}{{\sf Spin}} \nc{\Pin}{{\sf Pin}}
 
 \nc{\RR}{{\mathbb R}} \nc{\HH}{{\mathbb H}} \nc{\CC}{{\mathbb C}}
 \nc{\ZZ}{{\mathbb Z}} \nc{\FF}{{\mathbb F}} \nc{\NN}{{\mathbb N}}
 \nc{\GG}{{\mathbb G}} \nc{\JJ}{{\mathbb J}} \nc{\II}{{\mathbb I}}
 \nc{\KK}{{\mathbb K}} \nc{\DD}{{\mathbb D}}
 
 \nc{\ad}{\operatorname{ad}} \nc{\Ad}{\operatorname{Ad}}
 \nc{\coad}{\operatorname{coad}} \nc{\ct}{\operatorname{T}}
 \nc{\rank}{\operatorname{rank}} \nc{\Irr}{\operatorname{Irr}}
 \nc{\End}{\operatorname{End}} \nc{\Aut}{\operatorname{Aut}}
 \nc{\Inn}{\operatorname{Inn}} \nc{\Der}{\operatorname{Der}}
 
 \theoremstyle{plain}
 \newtheorem{thm}{Theorem}[section]
 \newtheorem{prop}[thm]{Proposition}
 \newtheorem{cor}[thm]{Corollary}

 \theoremstyle{definition}
 \newtheorem{defn}[thm]{Definition}
 
 \theoremstyle{remark}
 \newtheorem*{remark}{Remark}
 
 \newtheorem{exa}[thm]{Example}

\begin{document}
\title[The Adler Kostant Symes scheme in physics]
{The Adler Kostant Symes scheme in physics}

\author{Gabriela P. Ovando}
\address{G. Ovando: CONICET y ECEN-FCEIA, Universidad Nacional de Rosario, Pellegrini 250, 2000 Rosario, Santa Fe, Argentina}
\thanks{The author GO was partially supported by CONICET, ANPCyT and SECyT-UNC}

\email{ovando@mate.uncor.edu}


\begin{abstract} The purpose of this material is to review the Adler Kostant Symes scheme as a theory which can be developped succesfully in different contexts. It was useful to describe some mechanical systems, the so called generalized Toda, and now it was proved to be a tool  for the study of the linear approach to the motion of n uncoupled harmonic oscillators. The complete integrability of these systems has an algebraic description. In the original theory this is related to ad-invariant functions, but new examples show that new conditions should be investigated.
\end{abstract}

\thanks{{\it (2000) Mathematics Subject Classification}: 53C15, 53C55, 53D05, 22E25, 17B56 }

\maketitle

\section{Introduction}

In this work we are interested in the use of Lie theory to  understand  some Hamiltonian systems.  For the study of completely integrable systems one needs to identify the following: i) the symplectic structure, which gives the system its Hamiltonian character, ii) first integrals or constants of motion, iii) action angle variables, and the computation of their evolution. Indeed this is a very difficult approach but it is possible for systems related to certain Lie groups. 

To this end there are several methods with a common idea: the realization of canonical equations on Lie algebras, or on orbits of a certain action or on symmetric spaces. These ideas appeared in the 70's and were developped, under other by several authors as Adler, Fomenko, Kostant, Mischenko, Olshanetsy, Perelomov, Trofimov, Symes, etc. (see for instance \cite{Ad1} \cite{F-M1} \cite{F-M2} \cite{F-T} \cite{Ko1}  \cite{O-P} \cite{P} and \cite{Sy} and their references). 

In any method all of the above steps i) ii)  iii) are reflected by algebraic circumstances. One needs a way for imbedding a certain Hamiltonian system into a Lie algebra, effective methods for constructing sets of involution and the proof of the full integrability of a wide family of functions in involution.  

In this chapter we are concerned with so called  Adler-Kostant-Symes scheme, which brings together a mathematical framework with Lie theory but also consequences in the dynamics of the Hamiltonian system. This method 
 was successful when studying some mechanical systems such as the rigid body or the generalized 
 Toda lattice \cite{Ad2} \cite{Ko2} \cite{Sy} \cite{R2}. In this setting the phase space of the Hamiltonian systems become coadjoint orbits  represented on a Lie algebra 
 and  the functions in involutions are presented as ad-invariant functions. On the one hand  for this kind of functions, the corresponding Hamiltonian systems become a Lax 
 equation and on the other hand they are in involution on the orbits. Whenever studying Poisson commuting conditions the ad-invariance property can be replaced by a weaker one as in
 \cite{R1}. 
 In the framework of this theory what we
 need is a Lie algebra with an ad-invariant metric, a splitting of this Lie algebra into
 a direct sum as vector subspaces of two subalgebras and a given function. These algebraic tools were used with semisimple Lie algebras, where the Killing form is the natural candidate for the ad-invariant
 metric. 
 
 However there are more Lie algebras admitting an ad-invariant
 metric. We shall examplify  here how can be applied the theory for semisimple Lie algebras, and also for other ones, such as  the solvable ones. For the general case one should see that any Lie algebra with an ad-invariant metric  can be constructed by a double extension procedure, whose more simple application follows from $\RR^m$. In this way one gets  a solvable Lie algebra $\ggo$, that results a semidirect extension of the 2n+1-dimensional Heisenberg Lie algebra $\hh_n$ and that can be endowed with an ad-invariant metric which is an extension of a non degenerate  bilinear form on $\RR^{2n}$. But for other cases the resulting Lie algebras could be no semisimple and no solvable.

 In any case, the Lie algebra $\ggo$ splits naturally as  a direct sum of vector spaces of two subalgebras. Looking at the coadjoint orbits of one of the Lie subalgebras, one gets  Hamiltonian systems on these orbits and one can identify the original Hamiltonian system with one of these.  In particular for the restriction of the quadratic corresponding to the ad-invariant metric we obtain  a Hamiltonian system that  becomes a Lax equation, whose solution can be computed with the Adjoint representation. 
 
 As example we work out the Toda lattice and the linear equation of motion of n-uncoupled harmonic oscillators. The first one corresponds to a semisimple Lie algebra, and the second one is associated to a solvable one.  Furthermore it is proved that the Hamiltonian  for the last one is completely integrable on all maximal orbits. We notice that the functions in involution we are making use, are not ad-invariant and they do not satisfy the involution conditions of \cite{R1}.
 
 The setting for the second example applies for quadratic hamiltonians.   The Poisson commutativity conditions we get for some polynomials can be read off  in the Lie algebra $sp(n)$ of derivations of the Heisenberg Lie algebra of dimension 2n+1 $\hh_n$.    In  particular for the case of the motion of n-uncoupled harmonic oscillators we need a abelian subalgebra in the Lie algebra of isometries of the Heisenberg Lie group $\HH_n$, endowed with its canonical inner product. 
 
 This is not surprising if we consider that symplectic automorphisms of the Heisenberg Lie group produce symplectic symmetries of p-mechanical, quantum and classical dynamics  for more general systems than the linear ones (see  \cite{Ki2}).

The appearence of the Heisenberg Lie algebra related to the motion of n-uncoupled harmonic oscillators is not so surprising. In fact, it is known that in quantum mechanics a good approach to the simple harmonic oscillator is through the
 Heisenberg Lie algebra. In dimension three this is the Lie algebra generated by the
 position operator $Q$ = {\em multiplication by x}, the momentum operator $P = -i \frac
 {d}{dx}$ and $1$ with the only non trivial commutation relation
 $$[Q,P]=1$$
 These operators evolve according to the Heisenberg equations
 $$\frac{dP}{dt}=-Q \qquad \qquad \frac{dQ}{dt}= P$$
  
  An attempt to relate the classical mechanical system of the linear approximation of the motion of n-uncoupled harmonic oscillators was presented by the theory of p-mechanics, which  makes use of the representation theory of the Heisenberg Lie group to show that both quantum and classical mechanics can be derived from the same source (see for instance \cite{Ki1} \cite{Ki2}). This theory contructs a more general setting that unifies both quantum and classical mechanics. The starting point for p-mechanics is the method of orbit of Kirillov \cite{K1} \cite{K2}, which says that the orbits of the coadjoint representation of the Heisenberg Lie group parametrise all unitary irreducible representations \cite{F}. Thus non commutative representations are known to be connected with quantum mechanics. In the contrast commutative representations  are related to classical mechanics in the observation that the union of one dimensional representations naturally acts as the classical phase space in p-mechanics. In this theory the time evolution of both quantum and classical mechanics observables can be derived from the time evolution of  p-observables, choosen as particular functions or distributions on the Heisenberg Lie group. 
 
These considerations allow to suppose that new applications of the Adler Kostant Symes scheme are possible and maybe it comes a new time to understand old mechanical systems with  new tools, which should be developped for these purposes. As an introduction to the topic one can find exceptional ideas in the books of Arnold, Abraham and Marsden, Ratiu and Marsden, etc. all of them classics in the literature concerning classical mechanics.

The chapter is organised as follows: in the first part we present basic ideas concerning symplectic geometry. The second part is devoted to the Adler-Kostant-Symes scheme and the third part to the examples: on the one hand the Toda lattice with generalization in (\cite{Ko2} and \cite{Sy}), and on the other hand the systems corresponding to quadratic Hamiltonians on $\RR^{2n}$.

\section{Basic notions on symplectic manifolds}

In this section we present the basic elements to work with symplectic geometry. Some texts concerning this topic are \cite{L-M} \cite{CdS}.

\

Let $M$ denote a differentiable manifold. 

\begin{defn} A  2-form on $M$, $\omega$ is called a symplectic form if $d\omega =0$ and $\omega_p$ is non degenerate for every $p\in M$.

The pair $(M, \omega)$ is a symplectic manifold.
\end{defn}

It follows that the dimension of $M$ must be even. 

\begin{exa}\label{exa1}
 Let  $\RR^{2n}$ be the usual euclidean space equipped with global coordinates $x_1, \hdots, x_n$, $y_1, \hdots, y_n$. The 2-form given by  
$$\omega = \sum_{i=1}^n dy_i \wedge dx_i$$
defines a symplectic form on $\RR^{2n}$.

Note that if $(\,,\,)$ denotes the canonical inner product on $\RR^{2n}$ and $J$ the canonical complex structure 
$$J= \left( \begin{matrix} 0 & -I \\ I & 0\end{matrix} \right)$$
where $I$ is the identity $n \times n$ matrix, then 
$$\omega(X,Y) =(X,JY)$$
\end{exa}

\begin{exa} If $(M_1, \omega_1)$ and $(M_2, \omega_2)$ are symplectic manifolds, the direct product $M_1 \times M_2$ is a symplectic manifold.
\end{exa}

\begin{exa}  \label{coadac} Coadjoint orbits. Let 
$G$ denote a Lie group with Lie algebra $\ggo$ and let 
$\ggo^{\ast}$  be the dual space of $\ggo$. The 
 coadjoint action of $G$ on $\ggo^{\ast}$ is defined  as:
$$ g \cdot \varphi = \varphi \circ \Ad(g^{-1}) \qquad \qquad g\in G, \varphi\in \ggo^{\ast}.$$
Notice that the orbit throught $\varphi$ is the set $G\cdot \varphi =\{ g \cdot \varphi : g \in G\}$ and the isotropy subgroup at $\varphi$ is $G_{\varphi}=\{ g \in G \, : \, \varphi \circ Ad(g^{-1})=\varphi\}$; thus as usual one has $G \cdot \varphi= G / G_{\varphi}$.

The action of $G$ on $\ggo^{\ast}$ induces an action of $\ggo$ on $\ggo^{\ast}$ as 
$$X \cdot \varphi= - \varphi \circ  \ad(X) \qquad \qquad X\in \ggo, \varphi \in \ggo^{\ast}$$
that cames from the derivative
$$\frac{d}{dt}_{|_{t=0}} \exp(tX) \cdot \varphi = -\varphi \circ \ad(X),$$
in other words 
$$\tilde{X}(\varphi) = - \varphi \circ \ad(X)$$
is the infinitesimal generator induced by $X\in \ggo$ at $\varphi\in \ggo^*$. 
 
Any coadjoint orbit $G\cdot \varphi$ is a symplectic manifold with the 2-form 
 $$\omega_{\varphi}(\tilde{X},\tilde{Y})=-\varphi ([X,Y]), \qquad \varphi \in \ggo^{\ast}, \, X,Y\in \ggo.$$
called {\em the Kirillov-Kostant-Souriau} symplectic structure.
\end{exa}

Locally any symplectic manifold looks like the example (\ref{exa1}) above. This is a classical result of Darboux.

\begin{thm} {\bf Darboux.} Let $(M, \omega)$ be  a symplectic manifold. For every $p\in M$ there exists coordinate system $(U, (x^1, \hdots, x^n, y^1, \hdots, y^n))$ such that $p\in U$ and $\omega_{|_U}= \sum_{i=1}^n dy_i \wedge dx_i$.
\end{thm}

 The symplectic 2-form $\omega$ bilds a isomorphism on $T_xM$ for $x\in M$. In fact, since $\omega_x:T_x M \times T_xM$ is non degenerate, there is a linear isomorphism $K_x: T_x M \to T_x^{\ast}M$ defined by
 $$K_x(v) (u) = \omega_x(u,v)=(-i_v\omega)(u).$$
 
 Let $H:M \to \RR$ be a differentiable function, since its differential belongs to $T^{\ast}M$, via the isomorphism above we get a vector field $X_H$ given by
 \begin{equation}\label{hvf} X_H(x)=K_x( dH_x),
 \end{equation}  
that is, $X_H$ is the vector field on $M$ satisfying
$$v(H) = dH(v) = \omega(v, X_H)$$
and this is called the  {\em Hamiltonian vector field}   associated to the Hamiltonian function $H$.

\begin{exa} For the standard symplectic structure on $\RR^{2n}$ the isomorphism $K_x$ is given by
$K_x v= - J v$, where $J$ denotes the canonical complex structure on $\RR^{2n}$.

Let $H\in C^{\infty}(\RR^{2n})$, its associated Hamiltonian vector field is
$$X_H(m)=J(\nabla H) = \sum_i (\frac{\partial H}{\partial y_i} \frac{\partial}{\partial x_i}-\frac{\partial H}{\partial x_i} \frac{\partial}{\partial y_i}),$$
where $\nabla H$ is the gradient of $H$, with respect to the canonical inner product.
\end{exa}

\begin{defn} The Hamiltonian system for a Hamiltonian $H\in C^{\infty}(M, \omega)$ is 
\begin{equation} \label{hamsis}
x'(t)=X_H(x(t)).
\end{equation}
\end{defn}

\begin{exa} Let $H$ be a smooth function on $\RR^{2n}$, the  Hamiltonian equation is the classical one
$$\begin{array}{rcl}
x_i' & = & \frac{\partial f}{\partial y_i}\\
y_i' & = & - \frac{\partial f}{\partial x_i}
\end{array}
$$
where $x_i$ is actually $x_i(t)$, that is it depends on $t$, for all $i$ (and also for any $y_i$).
\end{exa}
 
\begin{exa} \label{exa2} On $\RR^{2n}$ a {\em quadratic Hamiltonian} is a smooth function as
$$H(x)=\frac12 (Ax,x)\qquad \mbox{ for } \qquad  A \mbox{ symmetric linear map,}$$
which yields the Hamiltonian system 
\begin{equation}\label{ham1}
x' = JA x
\end{equation}
 In classical mechanic this system describes ``small oscillations'', that is, it approximates the motion of a particle on $\RR^n$ or equivalently the motion of $n$ uncoupled particles on $\RR$, near an equilibrium position. 
 
 For instance  the motion of $n$-uncoupled harmonic oscillators near an equilibrium position can be approximated with $H$ a quadratic Hamiltonian as above  by taking $A=I$;   therefore (\ref{ham1}) becomes
 \begin{equation}\label{ho}
 \begin{array}{rcl}
 x_i'(t) & = & y_i(t)\\
 y_i'(t) & = & -x_i(t)
 \end{array}
 \end{equation}
 
 where $x(t)=(x_1(t), \hdots, x_n(t), y_1(t), \hdots, y_n(t))$.

In classical mechanics it is usual to name the coordinates as $x_i$ the position coordinates and $y_i$ as the velocity coordinates for every $i=1, \hdots, n$.
\end{exa}

\begin{defn} A diffeomorphism $\phi$ on a symplectic manifold $(M, \omega)$ is  symplectic if $\phi^{\ast} \omega = \omega$.
\end{defn}

Recall that the Lie derivative  on a smooth manifold $M$  given as
$$L_X T = \frac{d}{dt}_{|_{t=0}} \psi_t^{\ast}(T)$$
where $X$ is a vector field on $M$ with one parameter group $\psi_t$ and $T$ a tensor, satisfies the following identities
$$\begin{array}{rcl}
L_X & = & i_X d + d i_X\\
L_X i_Y & = & i_{L_X Y} + i_Y L_X =i_{[X,Y]} + i_Y L_X 
\end{array}
$$
For a proof see for instance \cite{Wa}.

\begin{defn} A vector field $X$ on a  symplectic manifold $(M, \omega)$ is symplectic if $L_X \omega=0$.
\end{defn}

\begin{prop} A vector field $X\in \chi(M)$ is symplectic if and only if the one parameter subgroup $\psi_t$ generated by $X$ is symplectic.
\end{prop}
\begin{proof} If $\psi_t$ is symplectic, using the definition of $L_X$ it is easy to see that $L_X \omega=0$. Conversely assume $L_X \omega =0$, then 
$$ \frac{d}{dt}_{|_{t=s}} \psi_t^{\ast} \omega =  \frac{d}{dt}_{|_{t=0}} \psi_t^{\ast} \psi_s^{\ast} \omega = L_X \psi_s^{\ast} \omega = \psi_s^{\ast} L_X \omega=0$$
hence $\psi_t^{\ast} \omega$ is constant. But $\psi_0=Id$ and so $\psi_t^{\ast}\omega = \omega$.
\end{proof}
 
 \begin{cor} i) If $\omega$ is a symplectic form then $L_X \omega= di_X \omega$.
 
 ii) A vector field on $(M, \omega)$ is symplectic if and only if $i_X \omega$ is closed.
 \end{cor}
 
 \begin{defn} A vector field $X$ on a symplectic manifold $(M,\omega)$ is Hamiltonian if and only if $-i_X\omega$ is exact.
 \end{defn}
 
 The vector field associated to a Hamiltonian function $H$ defined in (\ref{hvf}) is Hamiltonian. 
 
 Notice that   the fact of being $X$ Hamiltonian says that there is $H \in C^{\infty}(M)$ such that $dH =-i_X \omega$, therefore $X$ is symplectic. On the other hand for any $p \in M$ there always exists local solutions to $dH=-i_X \omega$ for any $X\in \chi(M)$. For global solutions we must ask extra conditions as below.
 
 \begin{prop} Let $(M, \omega)$ be a symplectic manifold such that  $H^1(M, \RR)=0$. Every symplectic vector field on $M$ is Hamiltonian.
 \end{prop}

A symplectic 2-form $\omega$ on a symplectic  manifold $M$ induces a  {\em Poisson} bracket $\{\,,\,\}$ on  $C^{\infty}M$ by:
$$\{f, g\}(p)=\omega_p(X_f, X_g)=X_f(g) =-X_g(f)\qquad
\qquad \mbox{ for any } f,g\in C^{\infty}M.$$

\begin{prop} Let $C^{\infty}(M)$ is a Lie algebra under the Poisson bracket defined above and $f \to X_f$ is a Lie algebra anti-homomorphism of $C^{\infty}(M)$ into $\chi(M)$.
\end{prop}
\begin{proof} Since $K_x:T_xM \to T_x^{\ast}M$ is a linear isomorphism, the map $f\to X_f$ is linear.  Now we  should prove that $[X_f, X_g]=X_{\{f,g\}}$. Using the properties of $L_X$  one gets
$$L_{X_f} i_{X_g} \omega = i_{[X_f, X_g]} \omega + i_{X_g} L_{X_f}\omega.$$
Since $L_{X_f}\omega=0$, one gets 
$$L_{X_f} i_{X_g} \omega = i_{[X_f, X_g]} \omega.$$
The Lie derivative on 1-forms follows
$$L_X \theta = i_X d \theta + di_X \theta.$$
Taking $i_{X_g}\omega=d g$ and applying above it holds
$$L_{X_f} i_{X_g} \omega = L_{X_f} dg  = i_{X_f} d^2 g + di_{X_f} dg = d(X_f(g))= d\{f,g\}.$$
Therefore $$i_{[X_f, X_g]}\omega = d\{f,g\}$$
Thus the left side of the equality above  $i_{[X_f, X_g]}\omega_x$ coincides with $-K_x(X_{\{f,g\}}$, and since $K_x$ is an isomorphism $[X_f, X_g]=-X_{\{f,g\}}$.
\end{proof}

\vskip 4pt

Recall that a {\em Poisson structure} is a bracket $\{\,,\,\}$ on a associative algebra $A$, such that 

$\bullet$ $\{\,,\,\}$ is a Lie bracket on $A$ and 

$\bullet$ $f\{g,h\} = \{fg, h\} + \{g, fh\}\qquad \mbox{ for all } f,g, h \in A.$

the last one is called the Leibnitz rule. In \cite{Sy} a such structure is called {\em Hamiltonian}.

\vskip 5pt

The space of smooth functions on a differentiable manifold is a associative Lie algebra, hence a natural space to be endowed with a Poisson structure.
The Proposition we already proved  says that whenever $(M, \omega)$ is a symplectic manifold,  $C^{\infty}(M)$ has  a Poisson structure induced by $\omega$: the Poisson bracket  $\{\, , \}$ is a Lie bracket and the Leibnitz rule holds, since any vector field is a derivation on $C^{\infty}(M)$.

\begin{exa}
On $\RR^{2n}$, the  Poisson structure associated to the standard symplectic form is given by
\begin{equation}\label{pb}
\{f, g \} = (\nabla f, J \nabla g)=  \sum_i \frac{\partial f}{\partial x_i}  \frac{\partial
 g}{\partial y_i}-
 \frac{\partial f}{\partial y_i} \frac{\partial g}{\partial x_i}.
\end{equation}
\end{exa}

\begin{exa} Let $\ggo$ be a Lie algebra and $\ggo^{\ast}$ its dual. As usual one identifies $\ggo^{\ast}$ with its tangent space. Given a function $F:\ggo^{\ast} \to \RR$, we define the gradient of $F$ at $\alpha\in \ggo^{\ast}$, denoted by $\nabla F(\alpha)$, as an element $\nabla F(\alpha)\in \ggo$ such that $\la \beta, \nabla F(\alpha) \ra = d F_{\alpha} (\beta)$ for any $\beta \in \ggo^{\ast}$, where $\la \, , \, \ra$ denotes the evaluation map. 

The Kirillov's Poisson bracket on $\ggo^{\ast}$ is given by
$$\{f,h\}(\alpha)=\la alpha, [\nabla f(\alpha), \nabla h(\alpha)]\ra.$$
 \end{exa}

\begin{prop} If $\{f,g\}=0$ then $g$ is constant on the integral curves of $X_f$.
\end{prop}
\begin{proof} Assume $x'(t)=X_f(x(t))$ then 
$$\frac{d}{dt} g(x(t)) = dg(x'(t)) = X_f(g)(x(t))=\{f,g\}(x(t))=0.
$$
\end{proof}

Thus $g$ is called a {\em constant of  motion} of the flow defined by $X_f$. Since $\{\, , \, \}$ is skew symmetric, $g$ is a constant of motion of $X_f$ if and only if $f$ is a constant of motion of $X_g$.
Constant of motion always exist, in fact $f$ is a constant of motion of $X_f$.

\begin{defn} A function $f$ on a 2n-dimensional Poisson manifold
$(M, \{ \,, \, \})$ is {\em completely integrable} if there exist $n$ functions $f_1, \hdots, f_n\in C^{\infty}M$ such that:

i) $\{f, f_i\}= 0$, $\{f_i, f_j\}=0$ for all $1\leq i, j \leq n$,

ii) The differentials $df_1, \hdots, df_n$ are linearly independent on 
a open set invariant under the flow of $X_f$.
\end{defn}

Two functions $f,g: M \to \RR$ such that $\{f,g\}=0$ are said to be in involution or Poisson commute.

 A subset $N \subset M$ is {\em invariant} under the flow of $X_f$ if the solution $x$ for the Hamiltonian system (\ref{hamsis}) corresponding to the Hamiltonian $f$ lies on $N$ if $x(0)\in N$.

\begin{exa}
On $\RR^{2n}$ for $H(x)=\frac12 (x,x)$ the polynomials
$$f_i(x)=\frac12 (p_i^2+q_i^2)\quad i=1, \hdots,n$$
shows that $H$ is completely integrable. In fact it is easy to check that $\{H, f_i\}=0=\{f_i, f_j\}$ for all $i=1, \hdots, n$.
 
 Let $F=(f_1, \hdots , f_n)$, then $F^{-1}(c)$ is a torus which is invariant under the flow generated by $X_H$. Let $(\theta_1, \hdots, \theta_n)$ denote the angle variable on the torus $F^{-1}(c)$. Then $(f_1, \hdots , f_n, \theta_1, \hdots, \theta_n)$ is a local coordinate on $\RR^{2n}$. With these coordinates, the Hamiltonian equation becames
 $$\begin{array}{rcl}
 f_i' & = & 0\\
 \theta_i' & = & -1
 \end{array}
 $$
and the coordinate functions satisfy $$\{f_i, f_j\}=\{\theta_i, \theta_j\} = 0, \qquad \{f_i, \theta_j\}=\delta_{ij},$$ 
 therefore the flow $X_h$ is linear on $F^{-1}(c)$ for $c\in \RR^n$.
 
 Moreover since the level sets $\{x\in \RR^{2n}: H(x)=c\}$ are compact we have action angle coordinates (see Liouville Theorem below).
\end{exa}

Generally $m$ Poisson commuting functions $f_1, f_2, \hdots, f_m$ on a symplectic manifold $(M, \omega)$ give rise to an action of $\RR^m$ on $M$. Let $(\psi_i)_t$ be the one parameter subgroup generated by $X_{f_i}$. Then 
 $$(t_1, \hdots, t_m)\cdot p=(\psi_1)_{t_1} (\psi_2)_{t_2} \hdots (\psi_m)_{t_m}(p).$$
 defines a $\RR^M$ action on $M$. Since $\{f_i, f_j\}=0$ for all $i,j$, the set $N=\{ x\in M,\,:f_i(x)=c_i\}$ is invariant under the $\RR^m$-action, for constants $c_1, \hdots, c_m$. If $N$ is compact, the $\RR^m$-action descends to a torus action on $N$. When $m=1/2 \dim M$, one gets the Liouville theorem.
 
 \begin{thm} [Liouville] Let $f$ be  a completely integrable function on $M$, with $\dim M=2n$, and assume $f_1:=f, f_2, \hdots, f_n$ are commuting Hamiltonians which are linearly independent and let $F=(f_1, \hdots, f_n):M \to \RR^n$ be proper. Then $F^{-1}(c)$ is invariant under the $\RR^n$ action and it descends to a torus $T^n$-action. Let $\theta_1, \hdots, \theta_n$ denote the angle coordinates on the invariant tori. Then $\{f_i, f_j\}=\{\theta_i, \theta_j\}=0$ and $\{f_i, \theta_j\}=c_{ij}(F)$ for some functions $c_{ij}:\RR^n\to \RR$. In particular, the flow of $X_f$ in coordinates $(f_1, \hdots, f_n, \theta_1, \hdots, \theta_n)$ is linear.
 \end{thm}

Coordinates as above, are called {\em action-angle variables} for the Hamiltonian system of $f$.

\section{Symplectic actions: the AKS-Scheme}

Let $M$ denote a differentiable manifold and let $G$ be a Lie group. An differentiable action of $G$ on $M$ is a differentiable map $\eta: G\times M \to M$, $\eta: (g,m) \to \eta(g,m):= g\cdot m$ such that
$$\begin{array}{lrlll}
\mbox{ i)} & e \cdot m & = & m &  \mbox{ for all } m\in M \mbox{ and }\\
\mbox{ii)} & (g h)\cdot m & = & g\cdot(h \cdot m) &\mbox{ for all } m\in M, g,h \in G.
\end{array}
$$
Notice that if $\eta$ is an action, the applications $\eta_g:M \to M$ given by $\eta_g(m)=g \cdot m$ are diffeomorphisms of $M$. In fact, $\eta_g$ are differentiable for any $g$ and they are diffeomorphisms since the inverse of any $\eta_g$ is $\eta_{g^{-1}}$ (see ii) above). Therefore an action of a Lie group on $M$ induces a representation of $G$ on $Diff(M)$ the diffeomorphisms of $M$, given by $g \to \eta_g$.

\begin{exa} Let $GL(n, \RR)$ denote the Lie group of non singular transformations of $\RR^n$. This acts on $\RR^n$ as evaluation: $A \cdot v= v$ for $A\in GL(n, \RR)$ and $v\in \RR^n$. It is easy to verify that this is in fact an action. 
\end{exa}

\begin{exa} Let $H$ be a Lie subgroup of a Lie group $G$, then  $H$ acts on $G$ by conjugation, $H\times G \to G$, $(h,x) = h^{-1} x h$, for any $h\in H$, $x\in G$. If $H$ is a normal subgroup, one can consider the action of $G$  on $H$ by conjugation.
\end{exa}

\begin{exa} Let $G$ be a Lie group with Lie algebra $\ggo$, then  $G$ acts on $\ggo$ by the Adjoint action,  $G\times \ggo \to \ggo$, $(g,X) = \Ad(g) X$, for any $g\in G$, $X\in \ggo$. Recall that $\Ad(g)= dI(g)_e$ where $I_g$ denotes the conjugation by $g$ (see the previous example). It is easy to see that $I_{gh} = I_g \circ I_h$ for all $g,h\in G$, hence the map $G \to GL(\ggo)$ is a representation of $G$, called the Adjoint representation. This has a correlative at the Lie algebra level, the adjoint representation: $\ggo \times \ggo \to \ggo$ given by $X \cdot Y = [X,Y]$ for all $X,Y\in \ggo$.
\end{exa}

Recall that in (\ref{coadac}) we defined the {\em coadjoint action} of a Lie group $G$ on the space $\ggo^{\ast}$, the dual of the Lie algebra 
$$g \cdot \varphi = \varphi \circ \Ad(g^{-1}) \qquad g\in G, \varphi \in \ggo^{\ast}$$
and also we gave the corresponding action of $\ggo$ on $\ggo^{\ast}$ by
$$X \cdot \varphi = -\varphi \circ \ad(X) \qquad X\in \ggo, \varphi \in \ggo^{\ast}.$$

The orbit of an action of a Lie group $G$ on a set $M$ is 
$$G \cdot m=\{ g \cdot m : g\in G\}$$
and the {\em isotropy}  or {\em stabilizer} group  of the action at the point $m$ is the closed subgroup of $G$ given by
$$G_m=\{ g\in G \mbox{ such that } g \cdot m=m\}.$$

It is known that the orbit  at $m$ is diffeomorphic to the quotient space of $G$ and the isotropy group, $G \cdot m \simeq  G/G_m$ (see \cite{Wa} for instance). Thus any curve at the orbit $G \cdot m$ through $m$ is 
$\gamma (t)=\exp tX \cdot m$ and this generates the infinitesimal vector $\tilde{X}$ at $T_m(G\cdot m)$ by
$$\tilde{X}(m)= \frac{\rm d}{\rm{dt}}_{|_{t=0}}  \exp tX \cdot m.$$
Hence the tangent space of a $G$-orbit at $m$ is 
$$T_m(G \cdot m)=\{ \tilde{X}, \quad X \in \ggo\}$$
being $\ggo$ the Lie algebra of $G$.

Assume $M$ and $N$ are two differentiable maps on which a given Lie group $G$ acts. A map $F: M \to N$ is called {\em equivariant} if $F( g \cdot m) = g \cdot F(m)$ for all $m\in M$, $g\in G$. The condition is also expressed as $F$ {\em intertwines} the two $G$-actions.

\begin{defn} Let $(M, \omega)$ be a symplectic manifold.  An action $\eta$ of a Lie group $G$ on $M$ is called symplectic if the diffeomorphisms $\eta_g$ are symplectic maps for any $g\in G$, that is $\eta_g^{\ast} \omega=\omega$.
\end{defn}

 The coadjoint orbits are examples of symplectic manifolds. Recall that they are endowed
 with the 2-form given by:
 $$\omega_{\beta}(\tilde{X},\tilde{Y})=-\beta ([X,Y]), \qquad \beta \in G\cdot \mu$$
 
 which is symplectic. In fact, it is closed since for  $X_1, X_2, X_3\in \ggo$ one has
 $$\omega([\tilde{X_1}, \tilde{X_2}], \tilde{X_3})=-\varphi([[X_1, X_2], X_3]]),$$
 hence
 $$d\omega(\tilde{X_1}, \tilde{X_2}, \tilde{X_3})=-\varphi([[X_1, X_2], X_3]])-\varphi([[X_2, X_3], X_1]])-\varphi([[X_3, X_1], X_2]])=0$$
 where the last equality holds after Jacobi for $[\cdot, \cdot]$. 
 
 The 2-form $\omega$ is non degenerate on a orbit: let $\varphi\in \ggo^{\ast}$ and let $X\in \ggo$ such that $\omega(\tilde{X}, \tilde{Y})=0$ for all $Y\in \ggo$. 
 
 Then $-\varphi([X,Y])=0$ for all $Y\in \ggo$, says that  $X \cdot \varphi=0$ implying that $X\in L(G_{\varphi})$. In fact $\exp tX\in G_{\varphi}$ if and only if $\exp tX \cdot \varphi = \varphi$ for $t$ near $0$. Thus taking derivative at $t=0$ we have $X \cdot \varphi=0$, and this is the set corresponding to the Lie algebra of $G_{\varphi}$. Since the tangent space of the orbit at $\varphi$  is $T_{\varphi} (G \cdot \varphi)=\ggo/L(G_{\varphi})$, one gets $\tilde{X}=0$.

  \begin{defn} An ad-invariant metric  on $\ggo$ is a bilinear map $\la \,,\,\ra: \ggo\times \ggo \to \RR$, which is   a non-degenerate symmetric and such that $\ad(X)$ is skew symmetric for any $X\in \ggo$, that is
 $$\la [X,Y], Z\ra +\la Y, [X,Z]\ra=0\qquad \qquad \mbox{ for all } X,Y, Z\in \ggo.$$
\end{defn}
 
  This ad-invariant metric gives rise to a bi-invariant pseudo Riemannian metric on a connected  Lie group $G$ with Lie algebra $\ggo$; bi-invariant means that the maps $\Ad(g)$ are isometries
for all $g\in G$, that is 
$$\la Ad(g) Y, Ad(g) Z\ra=\la Y, Z\ra \qquad \qquad \mbox{ for all } Y, Z\in \ggo, g\in G,$$
and conversely any bi-invariant pseudo Riemannian metric on $G$ induces an ad-invariant metric on its Lie algebra, just by taking derivative of the last equality at $t=0$ with  $g=\exp t X$.

\vskip 5pt

Examples of Lie algebras with ad-invariant metrics are:

a) semisimple Lie algebras with the Killing form;

b) semidirect products $\ggo \ltimes_{coad} \ggo^*$ with the canonical neutral metric
$$\la (x_1, \varphi_1),(x_2, \varphi_2)\ra=\varphi_1(x_2)+\varphi_2(x_1)$$

 An ad-invariant metric  $\la \,,\, \ra$ induces a diffeomorphism between the adjoint orbit $G\cdot X$ and the
coadjoint orbit $G\cdot \ell_X$ where $\ell_X (Y)=\la X, Y\ra$. In fact
$$ g \cdot \ell_X (Y)=\la X, Ad(g^{-1}) Y\ra= \la Ad(g) X, Y\ra \qquad \mbox{ for all } X,Y\in \ggo, g\in G, $$
implying that the map $\ell: X \to \ell_X$ is equivariant. Thus the adjoint orbits become symplectic manifolds with the 2-form:
$$\omega_X(\tilde{Y}, \tilde{Z})=\la X, [Y,Z]\ra\qquad \mbox{ for } X,Y, Z\in \ggo.$$

We shall consider these ideas to construct Hamiltonian systems on orbits that  are included on Lie algebras.

Recall that given a metric $\la \, ,\, \ra$ on $\ggo$  the gradient of a function $f:\ggo  \to \RR$ at the vector $X\in \ggo$ 
is defined  by
\begin{equation}\label{gra}
\la \nabla f(X), Y\ra = df_X(Y)\qquad \qquad Y\in \ggo. 
\end{equation}

\vskip 5pt

Suppose $\ggo_+$, $\ggo_-$ are Lie subalgebras of  the Lie algebra $\ggo$ such that
$$\ggo  = \ggo_+ \oplus \ggo_-$$ as a direct sum of linear subspaces, that is $(\ggo, \ggo_+, \ggo_-)$ is a product structure on $\ggo$. The Lie algebra $\ggo$ also splits as
$$\ggo=\ggo_+^{\perp} \oplus \ggo_-^{\perp},$$
 and  $$\ggo_{\pm}^{\perp}\quad \mbox{ is isomorphic as
vector spaces to }\quad \ggo_{\mp}^{\ast}.$$

This follows from the isomorphism $\ell:\ggo \to \ggo^{\ast}$. In fact, let $X\in \ggo_+^{\perp}$ maps to $\ell_X$. Since $\ell_X(Y)=0$ for all $Y\in \ggo_+$, the image of $\ell(\ggo_+^{\perp}$ belongs to $\ggo_-^{\ast}$, and the isomorphism follows from dimensions.

 Let $G_-$ denote a subgroup of $G$ with Lie algebra $\ggo_-$. Then the coadjoint action of
$G_-$ on $\ggo_-^{\ast}$ induces an action of $G_-$ on $\ggo_+^{\perp}$: for $g_-\in G_-$, $X\in \ggo_+^{\perp}$, $Y\in \ggo_-$ one has:
$$g_- \cdot \ell_X (Y)= \la X, \Ad(g^{-1}) Y\ra = \la \Ad(g) X, Y\ra= \pi_{\ggo_+^{\perp}}(\Ad(g_-)X ),Y\ra,$$
 where $\pi_{\ggo_+^{\perp}}$ denotes the projection of $\ggo$ on
 $\ggo_+^{\perp}$; therefore the action is given as
 $$ g_- \cdot X = \pi_{\ggo_+^{\perp}}(\Ad(g_-)X ),$$
 and $\ell:\ggo_+^{\perp} \to \ggo_-^{\ast}$ is equivariant.

  The infinitesimal generator corresponding to $Y_-\in \ggo_-$ is 
 $$\tilde{Y}_-(X)= \frac{\rm d}{\rm dt}_{|_{t=0}} \exp tY_- \cdot X=  \pi_{\ggo_+^{\perp}}([Y_-, X]) \qquad X\in \ggo_+^{\perp}.$$
 The orbit $G_-\cdot Y$ becomes a symplectic manifold with the
 symplectic structure  given by
 $$\omega_X(\tilde{U_-}, \tilde{V_-})=\la X, [U_-,V_-]\ra \qquad
 \mbox{ for } U_-, V_- \in \ggo_-, X\in G_- \cdot Y$$
 which is induced from  the Kostant-Kirillov-Souriau symplectic form on
 the coadjoint orbits in $\ggo_-^{\ast}$.

 Consider a smooth  function $f:\ggo \to \RR$ and restrict it to an orbit 
$G_-\cdot X 
 :=\mathcal M\subset \ggo_+^{\perp}$. Then the Hamiltonian vector field 
of the restriction 
 $H=f_{|_{\mathcal M}}$ is the infinitesimal generator corresponding to $-\nabla f_-$ , that is  \begin{equation}\label{e3}
 X_H(Y)=-\pi_{\ggo_+^{\perp}}([\nabla f_-(Y),Y])\end{equation}
 where  $Z_{\pm}$ denotes the projection of $Z \in \ggo$ with respect 
to the
decomposition $\ggo=\ggo_+\oplus \ggo_-$. In fact for $Y\in
 \ggo_+^{\perp}$, $V_-\in \ggo_-$ we have 
$$\begin{array}{rcl}
\omega_Y(\tilde{V_-}, X_H) & = & dH_Y(\tilde{V_-}) = \la \nabla f(Y),
\pi_{\ggo_+^{\perp}}([V_-,Y])\ra= \la \nabla f_-(Y),[V_-,Y]\ra \\
 & = & \la Y,[\nabla f_-(Y),V_-]\ra =  \omega_Y(\tilde{\nabla
   f_-(Y)},\tilde{V_-}).
\end{array}$$
 Since $\omega$ is non degenerate, one gets (\ref{e3}). 
 
 Therefore the 
Hamiltonian equation for $x:\RR \to \ggo$ 
 follows
 \begin{equation}\label{e4}
 x'(t)=-\pi_{\ggo_+^{\perp}}([\nabla f_-(x),x]).
 \end{equation}
In particular if $f$ is ad-invariant then $0 = [\nabla f(Y),Y]=
[\nabla f_-(Y),Y]+
[\nabla f_+(Y),Y]$. Since the metric is ad-invariant
$[\ggo_+, \ggo_+^{\perp}]\subset \ggo_+^{\perp}$, in fact
$$\la [\ggo_+, \ggo_+^{\perp}], \ggo_+ \ra= \la \ggo_+^{\perp}, [\ggo_+, \ggo_+]\ra =0.$$

 Hence the equation 
(\ref{e4}) takes the form
\begin{equation}\label{e5}
x'(t)=[\nabla f_+(x),x]=[x,\nabla f_-(x)],
\end{equation}
that is, (\ref{e4}) becomes a  {\em Lax equation}, that is, it can be written as $x'=[P(x),x]$.

If we assume now that the multiplication map $G_+\times G_- \to G$,
$(g_+, g_-) \to g_+g_-$,
is a diffeomorphism, then the initial value problem
\begin{equation}\label{e6}
\left\{ \begin{array}{rcl}
\frac{dx}{dt} & = & [\nabla f_+(x),x] \\
x(0) &  = & x_0
\end{array}
\right.
\end{equation}
can be solved by factorization. In fact if $\exp t\nabla 
f(x_0)=g_+(t)g_-(t)$, then
$x(t)=\Ad(g_+(t))x_0$ is the solution of (\ref{e6}).

{\sc Remark.} If the multiplication map $G_+ \times G_- \to G$ is a 
bijection
 onto an open subset of $G$, then equation (\ref{e4})  has a local solution in 
an interval
$(-\varepsilon, \varepsilon)$ for some $\varepsilon >0$.

\vskip 5pt

The theory we already exposed shows the application of Lie theory to the study of ODE's as in equation (\ref{e5}). Even when it is possible to give the solution, one need more information. This can be obtained from involution conditions. They help in some sense to control the solutions. 
 
A first step in the construction of action angle variables is to search for functions
which Poisson commute. 
 The  Adler-Kostant-Symes  Theorem shows a way to get functions in
 involution on the orbits $\mathcal M$.  We shall formulate it  in its classical Lie algebra
 setting.

 \begin{thm}[Adler-Kostant-Symes]\label{AKS1} Let $\ggo$ be a Lie algebra
 with an ad-invariant metric $\la \,, \,\ra$. Assume $\ggo_-, \ggo_+$  are Lie subalgebras
 such that $\ggo=\ggo_-\oplus \ggo_+$ as direct sum of vector  subspaces. Then any pair of
 ad-invariant functions on $\ggo$ Poisson commute on $\ggo_+^{\perp}$ (resp. on
 $\ggo_-^{\perp}$).
 \end{thm}
 Sometimes the ad-invariant condition is too strong, so the following version of the
 previous Theorem given by Ratiu \cite{R1} asks for a weaker condition.

 \begin{thm}\label{AKS2} Let $\ggo$ be a Lie algebra carrying  an ad-invariant metric $\la \,, \,\ra$. Assume it admits a splitting into a direct sum as vector spaces  $\ggo=\ggo_+\oplus  \ggo_-$, where $\ggo_+$ is an ideal
  and $\ggo_-$ is a Lie subalgebra. If $f,h$ are smooth Poisson commuting functions on
 $\ggo$, then the restrictions of $f$ and $h$ to $\ggo_+^{\perp}$ are in involution in
 the Poisson structure of $\ggo_+^{\perp}$.
 \end{thm}
\begin{remark} This theorem was used in \cite{R2} to prove the involution of the Manakov
integrals for the free n-dimensional rigid body motion.
\end{remark}

\section{Applications of the Adler-Kostant-Symes-scheme to classical mechanics}

In this section we show the explicit use of the theory above in some Lie groups and Lie algebras. The first example is done with semisimple Lie algebras, and it is known as the Toda Lattice. 

\subsection{The Toda lattice} The Toda lattice is the mechanical system which describes the motion of n particles on a line with an exponential restoring force, that is the Hamiltonian function on $\RR^{2n}$ is
$$H(x,y)=\frac12 \sum_{i=1}^n y_i^2+\sum_{i=1}^{n-1} e^{x_i-x_{i-1}}.$$
The phase space is $\RR^{2n}$ which is a symplectic manifold with its canonical symplectic structure. It follows that the Hamiltonian equation is

\begin{equation}\label{toda1}
\begin{array}{rcl}
x_k' & = & y_k \\
y_k' &  = & e^{x_{k-1} - x_k} - e^{x_k - x_{k+1}}
\end{array}
\end{equation}
and with  $e^{x_{0} - x_1} =0= e^{x_n - x_{n+1}}$. Flaschka considered a change of coordinates (called Falschka transform) as follows
$$\phi:\RR^{2n} \to \RR^{2n}, \qquad \phi(x,y)=(a,b)$$
where
$$
\begin{array}{rcll}
a_k & = &-\frac12 y_k & 1 \leq k \leq n,\\
b_k &  = & \frac12 e^{\frac12{x_k - x_{k+1}}} & 1 \leq k \leq n-1 \\
b_n & = & \frac12 e^{\frac{x_n}{2}}
\end{array}
$$
 Therefore the equation (\ref{toda1}) yields
 
$$
\begin{array}{rcll}
a_k' & = & 2(b_k^2 - b_{k-1}^2) & 1 \leq k \leq n \\
b_k' & = & b_k(a_{k+1} - a_{k}) & 1 \leq k \leq n 
\end{array}
$$
 
with $a_{n+1}=0=b_0$. Notice that $\sum_i y_i'=0$. Assume $\sum x_i=\sum y_i=0$ and let $V=\{(x,y) / \sum_i x_i = 0=\sum_i y_i\}$, then the system above becomes
\begin{equation}\label{toda2}
\begin{array}{rcll}
a_k' & = & 2(b_k^2 - b_{k-1}^2) & 1 \leq k \leq n-1 \\
b_k' & = & b_k(a_{k+1} - a_{k}) & 1 \leq k \leq n-1,
\end{array}
\end{equation}

 Consider  $\ggo$  the semisimple Lie algebra of  traceless real matrices $sl(n, \RR$ equipped with the ad-invariant metric $\la x, y\ra= tr(x,y)$ for all $x,y \in sl(n,\RR)$.

Let $\ggo_+=so(n)$ the Lie subalgebra of skew symmetric real matrices and $\ggo_-$ the Lie algebra of upper triangular matrices of trace zero. 

Then $\ggo_+^{\perp}$ is the space of real symmetric matrices in $sl(n,\RR)$ and $\ggo_-^{\perp})$ is the space of strictly upper triangular matrices in $sl(n, \RR)$.

The coadjoint orbit $\mathcal M= G_-\cdot x_0$ for $x_0=\sum_{i=1}^{n-1} e_{i, i+1} + e_{i+1,i}$ is the set  of tri-diagonal real symmetric matrices
$$\sum_{i=1}^n a_{i} E_{i,i}+ \sum_{i=1}^{n-1} b_i (E_{i, i+1} +  E_{i+1, i})\qquad \sum_i a_i=0,\quad  b_i >0 \quad \forall i.$$
where $E_{i,j}$ denotes the matrix with a 1 at the place $i,j$ and 0 in the others components.
 
Let $f: sl(n,R) \to \RR$ be the function given by $f(X)=\frac12 \la X, X\ra= \frac12 tr(X X)$. It is easy to see that the gradient of $f$ at $X$ is $X$, and hence applying the theory of the previous section we get  (\ref{e5})
$$ x'=[x_+ , x]$$
for $x_+ \subset \ggo_+$ a curve in $\ggo_+$. Writing the last system in terms of coordinates $(a,b)$ we get
the system (\ref{toda2}).

A generalization of this system can be read in \cite{Sy}, where also aplications of the theory to other differential equations are explained.

\subsection{The motion of n uncoupled Harmonic oscillators} Recall that the motion of $n$-uncoupled harmonic oscillators near an equilibrium position can be approximated with $H$ the quadratic Hamiltonian as $\frac12(x,x)$ where $(\, , \, )$ is the canonical inner product in $\RR^{2n}$. Let $\omega$ the canonical symplectic structure, the  corresponding Hamitonian system follows
 \begin{equation}\label{ho}
 \begin{array}{rcl}
 x_i'(t) & = & y_i(t)\\
 y_i'(t) & = & -x_i(t)
 \end{array}
 \end{equation}
 where $x(t)=(x_1(t), \hdots, x_n(t), y_1(t), \hdots, y_n(t))$.

 The associated Poisson structure on $\RR^{2n}$ is given as follows 
 \begin{equation}\label{pb}
\{f, g \} = (\nabla f, J \nabla g)=  \sum_i \frac{\partial f}{\partial x_i}  \frac{\partial
 g}{\partial y_i}-
 \frac{\partial f}{\partial y_i} \frac{\partial g}{\partial x_i}.
\end{equation}
for smooth functions
$f, g $ on $\RR^{2n}$. Thus with respect to the Lie bracket $\{ \,, \,\}$ the subspace over $\RR$
generated by the functions $H=\frac12\sum_i(x_i^2+y_i^2)$, the coordinates 
$x_i$, $y_i$, and $1$
form a solvable Lie algebra of dimension 2n+2, which is a semidirect extension
of the Heisenberg Lie algebra spanned by the functions $x_i,y_i,1$ i=1, $\hdots$,n. In fact 
they obey
the following non trivial rules
$$\{x_i, y_j \} =\delta_{ij} \qquad \{H, x_i \} =-y_i \qquad \{H, y_i\} =x_i.$$

In order to simplify notations let us rename these elements 
identifying $X_{n+1}$ with $H$, $X_i$ with $x_i$, $Y_i$ with $y_i$ and $X_0$
with the constant function 1
$$\begin{array}{rcl}
1 & \leftrightarrow & X_0\\
x_i & \leftrightarrow & X_i \\
y_i & \leftrightarrow & Y_i \\
H & \leftrightarrow & X_{n+1}
\end{array}
$$
 and set $\ggo$ denotes the Lie algebra
generated by these vectors with the Lie bracket $[\cdot , \cdot]$ derived from the
Poisson structure. This Lie algebra is known as a {\em oscillator} Lie algebra.

The Lie algebra $\ggo$ splits into a vector space direct sum
  $\ggo = \ggo_+ \oplus \ggo_-$, where $\ggo_{\pm}$ denote the Lie 
subalgebras
\begin{equation}
\label{deco} \ggo _- =  span\{X_0, X_i, Y_j\}_{i,j=1, \hdots n},\qquad 
 \qquad  \ggo_+ = \RR{X_{n+1}}.
\end{equation}

Notice that $\ggo_-$ is isomorphic to the 2n+1-dimensional Heisenberg Lie
 algebra  we denote $\hh_n$.

The quadratic form on $\ggo$ which for
 $X =x_0(X) X_0  + \sum_i (x_i(X) X_i + y_i(X) Y_i) + x_{n+1}(X)X_{n+1}$ is given by
$$f(X)=  \frac12  \sum_i(x_i^2+ y_i^2)+ x_0x_{n+1}$$
induces an ad-invariant metric on $\ggo$ denoted by $\la \,, \,\ra$. 
It is easy to show that the gradient of $f$ at a point $X$ is  $$ \nabla f(X) = X.$$

The restriction of the quadratic form to $\vv:=span\{X_i, Y_j\}$ i, j=1, $\hdots $,
n, coincides with the canonical one $(\, , \, )$ on $\RR^{2n}\simeq \vv$.

 The  metric induces a decomposition of  the Lie
 algebra $\ggo$
 into a  vector subspace direct sum of
$\ggo_+^{\perp}$ and $\ggo_-^{\perp}$ where $$ \ggo_-^{\perp} =
span\{X_0\} \qquad \qquad \ggo_+^{\perp} = \RR X_{n+1} \oplus \,span\{X_i,
Y_j\}_{i,j=1, \hdots, n},$$
and it also induces linear isomorphisms
$\ggo_{\pm}^{\ast}\simeq\ggo^{\perp}_{\mp}$. Let $G$ denote a Lie
group with Lie algebra $\ggo$ and $G_{\pm}\subset G$ is a Lie subgroup
whose Lie algebra is $\ggo_{\pm}$. Hence the  Lie subgroup
$G_-$ acts on $\ggo_+^{\perp}$ by the ``coadjoint'' representation; which in 
terms of
$U_-\in \ggo_-$ and $V\in \ggo_+^{\perp}$ is given by
\begin{equation}\label{m2}
\begin{array}{rcl}
\ad^{\ast}_{U_-} V  & =  &  x_{n+1}(V) \sum_i(y_i(U)
X_i - x_i(U) Y_i)
\end{array}
\end{equation}
It is not difficult to see that the orbits are 2n-dimensional
if $x_{n+1}(V) \ne 0$ and furthermore $V$ and $W$ belong to the same orbit if and only if
$x_{n+1}(V)=x_{n+1}(W)$, hence the orbits are parametrized by the $x_{n+1}$-coordinate;  
 so we denote them by $\mathcal M_{x_{n+1}}$. They are topologically
like $\RR^{2n}$. In fact $\mathcal M_{x_{n+1}}= G_- \cdot V\simeq
\HH_n/Z(\HH_n)$, where $\HH_n$ denotes the Heisenberg Lie group with center
 $Z(\HH_n)$.
 
 Equipp these  coadjoint orbits  with the canonical symplectic structure, that is for $U_-, V_- \in \ggo_-$ take 
 $$\omega_Y(\tilde{U}_-, \tilde{V}_-)=\la Y, [U_-, V_-]\ra = x_{n+1}(Y) \sum_{i=1}^n (x_i(U_-) y_i(V_-)-x_i(V_-)y_i(U_-)).$$
Indeed on  the orbit $\mathcal M_1$  the coordinates $x_i, y_j$, $i,j =1, \hdots n$, are   the canonical 
symplectic coordinates  and one can identify this orbit with $\RR^{2n}$ in a natural way. This says that the identification is a symplectomorphism between $\RR^{2n}$ with the canonical symplectic structure and the orbit with the Kirillov-Kostant-Souriau symplectic form.
 
 Consider $H$, the 
 restriction to a orbit $\mathcal M_{x_{n+1}}$ of  the function $f$. Since $f$ is ad-invariant the
Hamiltonian system of $H=f_{|_{\mathcal M_{x_{n+1}}}}$ reduces to
\begin{equation}\label{osc}
\begin{array}{rcl}
\frac{\rm dx}{\rm dt}& = & [x_{n+1}X_{n+1}, x_{\vv}+ x_{n+1} X_{n+1}]\\
 x(0) & = & x^0
\end{array}
\end{equation}
where $x^0= x_{\vv}^0 +x_{n+1}^0 X_{0}$ and $x_{\vv}^0=\sum_i (x_i^0 X_i + y_i^0
Y_i)$.

For $x_{n+1}\equiv x_{n+1}^0\equiv 1$ this system is that one we get on $\RR^{2n}$.

The trajectories $x(t)$ with coordinates $x_i(t)$, $y_j(t)$, $x_{n+1}^0$  are parametrized 
circles of angular velocity $x_{n+1}^0$,  for all i,j, that is

  $$
\begin{array}{rcl}
x_i(t) & = &  x_i^0 \cos(x_{n+1}^0 t)  + y_i^0 \sin( x_{n+1}^0 t)\\
y_j(t) & = & - x_j^0 \sin(x_{n+1}^0 t) + y_j^0 \cos(x_{n+1}^0 t)\\
x_{n+1}(t) & = &  x_{n+1}^0
\end{array}
$$
 This solution coincides
with that computed in the previous section, when we considered systems 
on coadjoint orbits. In fact it can be written as 
$$x(t)= \Ad(exp\,\, t x_{n+1}^0 X_{n+1}) x^0,$$
and one verifies that the  flow at the point $X^0 \in \ggo_+^{\perp}$ is 
\begin{equation}\label{flow}
\begin{array}{rcl}
\Delta^t(X^0) & = & 
\sum_i [(x_i^0 \cos(x_{n+1}^0 t) + y_i^0 \sin( x_{n+1}^0 t))X_i + 
(-x_i^0 \sin(x_{n+1}^0 t) + \\ \\
& & y_i^0 \cos(x_{n+1}^0 t))Y_i] +   x_{n+1}^0 X_{n+1}
\end{array}
\end{equation}

By taking $L$ and $M$ the following matrices:
$$ M = \left(
\begin{matrix}
0 & x_{n+1} & 0& 0 &&& &0 & 0\\ -x_{n+1} & 0 & 0 & 0 
&&&&0 &
0 \\ 0 & 0 & 0 & x_{n+1} &&&&0& 0\\ 0& 0 & -x_{n+1}&
0&&&&0&0\\ & & & & \ddots & & &\vdots & \vdots\\ & & & & & 0 &
x_{n+1} &0 & 0\\ & & & & 0& -x_{n+1}& 0 &0& 0\\ 0& 0& 
\hdots
& & & & & 0 & 0\\ 0 & 0 & \hdots & & & &  & 0 & 0
\end{matrix}
\right) $$
$$ L = \left(
\begin{matrix}
0 & x_{n+1} & 0& 0 & & & & &x_1\\ -x_{n+1} & 0 & 0 & 0  
& &
& & & y_1\\ 0 & 0 & 0 & x_{n+1} & & & & & x_2\\ 0& 0 &
-x_{n+1} & 0 & & & & & y_2\\ & & & & \ddots & &  & \vdots &
\vdots\\ & & & & & &  x_{n+1} & 0& x_n\\ & & & & &  -x_{n+1}
 & 0 &0 & y_n\\ -\frac12 y_1& \frac12 x_1&-\frac12 y_2 &\frac12
x_2 &\hdots & -\frac12 y_n& \frac12 x_n &  0 & 0\\ 0& 0 & 0 & 0 &
\hdots  & 0& 0& 0 & 0
\end{matrix}
\right) $$
we get  $L^{\prime} = [M, L]= ML -
LM$, the Lax pair equation.

\section{Quadratic Hamiltonians and coadjoint orbits} In this section we shall prove that Hamiltonian systems corresponding to quadratic Hamiltonians in $\RR^{2n}$  of the form $H(x) = \frac12(Ax,x)$ where $A$ is a non singular symmetric map, can be described using the  scheme of  Adler-Kostant-Symes  on a solvable Lie algebra. 

Let us consider the linear system of one 
degree  of freedom on $\RR^{2n}$ with Hamiltonian given by:
$$
H(x)= \frac12 (Ax, x)
$$
where   $x=(q_1, \hdots, q_n, p_1, \hdots, p_n)$ is a vector in $\RR^{2n}$ written
  in a symplectic
 basis and $A$ is a non singular symmetric linear operator with respect to the canonical inner product
 $(\, ,\, )$. This yields the following  Hamiltonian  equation

 $$(\ref{ham1}) \qquad \qquad \qquad \qquad \qquad 
 x'=JA x , \qquad\qquad \mbox{ with } J = \left( \begin{matrix} 0 & -Id \\ Id & 0
 \end{matrix} \right)
 $$
  
\noindent and being $Id$  the identity.  
The phase space for this system is $\RR^{2n}$.
We shall construct a solvable Lie algebra that admits an ad-invariant metric on which the system (\ref{ham1}) can be realized as a Hamiltonian system on coadjoint orbits. Moreover it can be written as a Lax pair equation.

Let $b$ denote the non degenerate bilinear form on $\RR^{2n}=span \{X_i, Y_j\}_{i,j=1}^n$ given by $b(X,Y)=(AX,Y)$.  In our terms, $b$ defines a metric on $\RR^{2n}$ but it is not necessary definite. Note that  the linear $JA$ is non singular and skew symmetric with respect to $b$, where $J$ is the canonical complex structure on $\vv\simeq \RR^{2n}$ as above:
$$b(JAX,Y)=(AJAX,Y)=(JAX,AY)=-(AX,JA Y)=-b(X,JA).$$
Let $\ggo$ denote the Lie algebra $\ggo$ which as vector space is the diract sum 
 $$\ggo=\RR X_0  \oplus \vv \oplus \RR X_{n+1}$$ 
 where $\vv= \RR^{2n}$ and with the Lie bracket given by the non trivial relations
\begin{equation}\label{lb}
[U, V] = b(JA U, V) X_0 \qquad \ [X_{n+1}, U ] = JA U \quad\mbox{ for all } U\in \vv. 
\end{equation}

Thus in this way one defines a structure of a solvable Lie algebra on $\ggo$. Note that $A=Id$ is the particular case we considered in the previous subsection. 

This Lie algebra $\ggo$  can be equipped with the ad-invariant metric defined by
\begin{equation}\label{metric}
\la x_0^1X_0 + U^1 + x_{n+1}^1 X_{n+1}, x_0^2X_0 + U^2 + x_{n+1}^2 X_{n+1}\ra = b(U^1, U^2) + ( x_0^1 x_{n+1}^2 
+ x_0^2 x_{n+1}^1).
\end{equation}

Thus if  $\la \,, \,\ra_{\vv}$ denotes the restriction of the metric of $\ggo$ to $\vv=span\{X_i, Y_j\}_{i,j=1, \hdots , n}$, then clearly $\la \, , \,\ra$  is a generalization of the non degenerate symmetric bilinear map $b$  of $\RR^{2n}$. Moreover  $\ggo$ admits  a orthogonal splitting 
$$ \ggo=   span\{X_0,X_{n+1}\}\oplus \vv.$$ 

Denote by  $\ggo_{\pm}$ the Lie subalgebras 
$$\ggo_+=\RR X_{n+1}, \qquad \ggo_-=\RR X_0 \oplus span\{X_i,, Y_i\}.$$
They induce the  splitting of $\ggo$ into a vector space direct sum 
  $\ggo = \ggo_+ \oplus \ggo_-$, which by  the ad-invariant metric gives  the following linear decomposition $\ggo= \ggo_+^{\perp} \oplus \ggo_-^{\perp}$, direct sum as vector spaces, for 
  $$ \ggo_-^{\perp} = \RR X_0 \qquad \qquad \ggo_+^{\perp} = span\{X_i, Y_i\}_{i=1, \hdots, n}
\oplus \RR X_{n+1}.$$
   Note that  $\ggo_-$ is  an ideal of $\ggo$ isomorphic to the 2n+1-dimensional
Heisenberg Lie  algebra $\hh_n$.

Let $G$ denote a Lie group with Lie algebra $\ggo$, set $G_-\subset G$
the Lie subgroup with Lie subalgebra $\ggo_-$. As we already explained  $G_-$ acts on  $\ggo_+^{\perp}$ by the coadjoint
action  $$g_- \cdot X= \pi_{\ggo_+^{\perp}}(\Ad(g_-)X ) \quad g_-\in G_-, \quad X \in
\ggo_+^{\perp},$$
 where $\pi_{\ggo_+^{\perp}}$ is the projection of $\ggo$ on
 $\ggo_+^{\perp}$, which 
 in infinitesimal terms gives the following  action of $\ggo_-$ on $\ggo_+^{\perp}$ 
\begin{equation}\label{m22}
\begin{array}{rcl}
\ad^{\ast}_{U} V : = U \cdot V   & =  &   x_{n+1}(V) JA X_{\vv}(U)
\end{array} \qquad \mbox{ for } U\in \ggo_-, \, V\in \ggo_+^{\perp}.
\end{equation}
being $X_{\vv}(U)$ the projection of $U$ onto $\vv$ with respect to the orthogonal splitting 
$\ggo = span\{X_0,\, X_{n+1}\} \oplus \vv$. 

The orbits are 2n-dimensional if $x_{n+1}(V) \ne 0$  and furthermore $V$ and $W$ belong to the same orbit if and only if
$x_{n+1}(V)=x_{n+1}(W)$, and therefore one  parametrizes the orbits by the $x_{n+1}$-coordinate  and one enotes  them by $\mathcal M_{x_{n+1}}$. The orbits are topologically
like $\RR^{2n}$ since they are 
diffeomorphic to the quotient $\HH_n/Z(\HH_n)$, if  $Z(\HH_n) = \RR X_0$ is the center of the Heisenberg subgroup.

Endow the orbits with the canonical symplectic structure of the coadjoint
orbits, that is for
$X\in \ggo_+^{\perp}$, $U_-, V_-\in \ggo_-$ set
$$\omega_X(\tilde{U_-}, \tilde{V_-})=\la X, [U_-, V_-]\ra= x_{n+1}(X) b(JA U_{\vv}, V_{\vv}).$$

Consider $f:\ggo \to \RR$ the ad-invariant function given by 
$$f(X)= \frac12 \la X, X\ra.$$
 The gradient of the function $f$ at a point $X$  is the so called position vector 
$$\nabla f(X) = X.$$ Since $f$ is
ad-invariant the 
Hamiltonian system of $H=f_{|_{\mathcal M_{x_{n+1}}}}$, the restriction of $f$  
 to the orbit $\mathcal M_{x_{n+1}}$,  given by (\ref{e5}) becomes
\begin{equation}\label{ipe}
\begin{array}{rcl}
\frac{dx}{dt}& = & [\nabla f_+(x),x] = [x_{n+1}X_{n+1},  x_{\vv} + x_{n+1} X_{n+1}]= x_{n+1} JA x_{\vv}\\
 x(0) & = & X^0
\end{array}
\end{equation}
where $X^0 \in \ggo_+^{\perp}$. 

Thus this  Hamiltonian system written as a Lax pair equation is equivalent to  (\ref{ham1}) for $x_{n+1} = x_{n+1}^0 = 1$.  The solution $X(t)$ for the initial condition  $X^0 \in \ggo_+^{\perp}$ can be computed via  the Adjoint map on $G$, that is,
$$X(t)= \Ad(exp\,\, t x_{n+1}^0 X_{n+1}) X^0.$$
The previous explanations prove the following result.

\begin{thm}[\cite{O2}] Let $H(X)= \frac12(AX,X)$ be a quadratic Hamiltonian on $\RR^{2n}$ with corresponding Hamiltonian system (\ref{ham1}). Then $H$ can be extended to a quadratic function $f$ on a solvable Lie algebra $\ggo$ containing the Heisenberg Lie algebra as a proper ideal. The function $f$ induces  a Hamiltonian system on coadjoint orbits of the Heisenberg Lie group, that can be written as a  Lax pair equation and which is equivalent to (\ref{ham1}). Moreover the trajectories on $\RR^{2n}$ for the initial condition $V^0$ can be computed with help of the Adjoint map on $\ggo$. Explicitely  they are the curves $x(t) = \exp^{t J A} V^0$, where $\exp$ denotes the usual exponential map  of matrices.
\end{thm}

If we take $L, M \in M(2n+2, \RR)$ as 
$$ M = \left(
\begin{matrix}
x_{n+1}JA & 0 & z \\ 
0 & 0 & 0 \\ 0 & 0 & 0
\end{matrix}
\right) \quad 
L = \left(
\begin{matrix}
x_{n+1} JA & 0 & z \\ 
 i\frac12z^T & 0 & 0 \\ 0 & 0 & 0 \end{matrix}
\right)
$$
where $z^T=(x_1,  x_2, \cdots, x_n, y_1, y_2, \hdots, y_n)$ 
then the Hamiltonian equation can be written in the following way
$$L^{\prime} = [M, L].$$

\begin{exa}[The motion of n-uncoupled inverse pendula] As example of the previous construction consider  the linear approximation 
of the motion of n uncoupled inverse pendula. This corresponds to the Hamiltonian $
H(x)= \frac12 (Ax, x)$ with  
$$A=\left( \begin{matrix}  Id & 0 \\ 0 & -Id
\end{matrix} \right).
$$
This yields  the Hamiltonian system $x' = JAx$, which in coordinates takes the form
\begin{equation}\label{33}
\begin{array}{rclcl}
\frac{dx_i}{dt} &  = & y_i\\
 \frac{dy_i}{dt} & = & {x_i}
\end{array}
\end{equation}
 
As we said the phase space  is $\RR^{2n}$. In  the setting of the AKS scheme we can construct  coadjoint orbits $\mathcal M$ of the Heisenberg Lie group,  that are included in a solvable Lie algebra $\ggo$ with Lie bracket (\ref{lb}) and ad-invariant metric (\ref{metric}). The Hamiltonian system for the restriction to the orbits of the ad-invariant function on $\ggo$, $f(X)=\frac12\la X, X \ra$,  can be written as 
\begin{equation}\label{ipel}
\begin{array}{rcl}
\frac{dx}{dt}& = & [x_{n+1}X_{n+1},  x_{\vv} + x_{n+1} X_{n+1}]\\
 x(0) & = & X^0
\end{array}
\end{equation}
where $X^0= \sum_i (x_i^0 X_i + y_i^0 Y_i)+x_{n+1}^0 X_{n+1}$. The Hamiltonian system above on the coadjoint orbit $\mathcal M_1$  written in coordinates is clearly equivalent to (\ref{33}). 

The trajectories on $\ggo_+^{\perp}$, $x= \sum_i (x_i(t)X_i+ y_i(t) Y_i) + x_{n+1} X_{n+1}$ are 
parametrized  by
  $$
\begin{array}{rcl}
x_i(t) & = &  x_i^0 \cosh(x_{n+1}^0 t) + y_i^0 \sinh( x_{n+1}^0 t)\\
y_i(t) & = &  x_i^0 \sinh(x_{n+1}^0 t) + y^0_i \cosh(x_{n+1}^0 t)\\
x_{n+1}(t) & = &  x_{n+1}^0
\end{array}
$$

The flow at the point $X^0 \in \ggo_+^{\perp}$ is 
\begin{equation}\label{iflow}
\begin{array}{rcl}
\Delta^t(X^0) & = & \sum_i[( x_i^0 \cosh(x_{n+1}^0 t) - y_i^0 \sinh( x_{n+1}^0 t)X_i
 + \\
 & & + (x_i^0 \sinh(x_{n+1}^0 t) + y_i^0 \cosh(x_{n+1}^0 t)Y_i] +  
  x_{n+1}^0 X_{n+1}
  \end{array}
\end{equation}

\vskip .2cm

The system (\ref{ipel}) is a Lax pair equation $L^{\prime} = [M, L]= ML -
LM$, and has a matricial representation by choosing $L$ and $M$ the following matrices in $M(2n+2,\RR)$:
$$ M = \left(
\begin{matrix}
0 & x_{n+1}& 0& 0 &&& &0 & 0\\ x_{n+1}& 0 & 0 & 0 
&&&&0 &
0 \\ 0 & 0 & 0 & x_{n+1} &&&&0& 0\\ 0& 0 & x_{n+1}&
0&&&&0&0\\ & & & & \ddots & & &\vdots & \vdots\\ & & & & & 0 &
x_{n+1}&0 & 0\\ & & & & 0& x_{n+1}& 0 &0& 0\\ 0& 0& 
\hdots
& & & & & 0 & 0\\ 0 & 0 & \hdots & & & &  & 0 & 0
\end{matrix}
\right) $$
$$ L = \left(
\begin{matrix}
0 & x_{n+1} & 0& 0 & & & & &x_1\\ x_{n+1} & 0 & 0 & 0  
& &
& & & y_1\\ 0 & 0 & 0 & x_{n+1} & & & & & x_2\\ 0& 0 &
x_{n+1} & 0 & & & & & y_2\\ & & & & \ddots & &  & \vdots &
\vdots\\ & & & & & &  x_{n+1} & 0& x_n\\ & & & & &  x_{n+1}
 & 0 &0 & y_n\\ -\frac12 y_1& \frac12 x_1&-\frac12 y_2 & \frac12
x_2 &\hdots & -\frac12 y_n& \frac12 x_n &  0 & 0\\ 0& 0 & 0 & 0 &
\hdots  & 0& 0& 0 & 0
\end{matrix}
\right) $$
\end{exa}

Now we shall investigate involution conditions on the coadjoint orbits of the Heisenberg Lie group for the restrictions of the quadratic functions $f(X)=\frac12 \la X,X\ra$, where $\la \,,\, \ra$ denotes the ad-invariant metric on the solvable Lie algebra $\ggo$. 

Let $g_i, g_j$ be two quadratics on $\RR^{2n}$ that are  realted to the symmetric 
maps $A_i, A_j:\vv \to \vv$ respectively, that is
$$g_i(X) = \frac12 (A_i X, X) \qquad \qquad g_j(X)=\frac12 (A_jX,X).$$ 
Consider quadratic functions on the solvable Lie algebra $\ggo$, which are extensions of $g_i, g_j$ to $\RR X_0 \oplus \RR X_{n+1}$, for instance as 
$$g_i(X) = \frac12 (A_i X_{\vv}, X_{\vv}) + x_0 x_{n+1}\qquad \qquad g_j(X)=\frac12 (A_jX_{\vv},X_{\vv}) +x_0 x_{n+1}.$$
For the following results these extensions are not unique. For instance extending them trivially we get the same conclusions.

 Let $H_i, H_j$ denote the restrictions
of $g_i, g_j$ to the orbits $\mathcal M_{x_{n+1}}$ and let  $X\in \mathcal M_{x_{n+1}} \subset \ggo_+^{\perp}$. The symplectic structure on the orbits induces a Poisson bracket which for 
the functions $H_i, H_j$ follows:  
$$
\{H_i, H_j\}(X)  =  \la X, [\nabla {g_i}_-(X), \nabla {g_j}_-(X)]\ra 
$$
By computing one can see that the gradients of $g_i$ and $g_j$ are
$$\nabla g_i(X) = A^{-1} A_i X_{\vv}+x_0X_0 + x_{n+1} X_{n+1} \qquad \nabla g_j(X) = A^{-1} A_j X_{\vv}+x_0 X_0 + x_{n+1} X_{n+1}.$$

Thus we are ready to prove the following result. 

\begin{thm} [\cite{O2}] \label{c11} The functions $H_i, H_j$ are in involution on the orbits $\mathcal
M_{x_{n+1}}$ if and only if 
\begin{equation}\label{c1}
[JA_i, JA_j]=0
\end{equation}
where $J$ is the canonical complex structure on $\RR^{2n}$.
\end{thm}
\begin{proof} Let $X\in \mathcal M_{x_{n+1}} \subset \ggo_+^{\perp}$. For
the functions $H_i, H_j$ the Poisson bracket on the orbit $\mathcal M_{x_{n+1}}$ follows:
$$\begin{array}{rcl}
\{H_i, H_j\}(X) & = & \la X, [A_i X_{\vv}, A_jX_{\vv}]\ra = \la x_{n+1} [X_{n+1},
A^{-1}A_i X_{\vv}], A^{-1}A_j X_{\vv}\ra \\
& = &  x_{n+1} \la J A_i X_{\vv}, A^{-1} A_jX_{\vv}\ra = x_{n+1} (J A_i X_{\vv}, A_jX_{\vv})
\end{array}
$$
  Therefore $\{H_i, H_j \}(X)=0$ if and only if
$( A_j J A_i X_{\vv}, X_{\vv}\ra=0$ which is equivalent to  
$A_j J  A_i= A_i J  A_j$, if and only if $JA_j J  A_i= JA_i J  A_j$, that is $[JA_i, JA_j]=0$.
\end{proof}

The natural question is what is the meanning of (\ref{c1})?

Fix $\la \, , \,\ra'$ the inner product on $\hh_n$ defined so  that the
vectors $X_i, Y_j, X_0$ are orthonormal for all i,j=1,$\hdots$, n. The  metric is an extension of the canonical one on $\RR^{2n}$. The Lie bracket on $\hh_n = \RR X_0 \oplus \vv$ where $\RR^{2n}\simeq \vv = span\{ X_i, Y_j\}_{i,j=1, \hdots, n}$ is expressed as 
$$\la [X, Y], x_0 X_0 \ra' = x_0 \la J X, Y \ra' \quad \mbox{ with } J \mbox{ as in } (\ref{ham1})$$
and note that $\la \,, \,\ra_{|_{\vv \times\vv}} = (\, ,\,)$. 
A derivation $D$ of $\hh_n$ acting trivially on the center must satisfy $[DU, V]= - [U, DV]$ for all $U, V \in \vv$. Equivalently in terms of $\la \,,\, \ra'$, we have that a map $D$ in $\hh_n$ is a derivation acting trivially on the center of $\hh_n$ if and only if the restriction of $D$ to $\vv$ (denoted also $D$) satisfies
$$ ( J DU, V ) = - (JU, DV) \qquad \mbox{ for all } U, V \in \vv,$$ 
where we replaced $\la \,, \,\ra'_{\vv}$ by $(\, ,\, )$ since they coincide on $\vv \simeq \RR^{2n}$. Denote by $\dd$  the set of derivations on $\hh_n$ acting trivially on the center of $\hh_n$. 

\begin{thm} \label{bij} There is a bijection between the set of derivations of $\hh_n$ acting trivially on the center and the set $\mathfrak{so}(n)$ of symmetric linear maps on $\RR^{2n}$. This correspondence is given by  $D \in \dd \to JD \in \mathfrak{so}(n)$, where $J$ is the complex structure as in (\ref{ham1}). 
\end{thm}

\begin{cor} If there exists an n-dimensional abelian subalgebra on $z(JA)_{\dd}$, where 
$$z(JA)_{\dd}=\{ D \in \dd \mbox{ such that } [D, JA]=0\}$$
 then the Hamiltonian function $H$ restriction of the function $f(X)= \frac12(AX,X)$ is completely integrable on the orbits $\mathcal M_{x_{n+1}}$ for $x_{n+1} \neq 0$.
\end{cor}
\begin{proof} The previous theorem says that  the restrictions to the orbit $\mathcal M_{x_{n+1}}$ of the functions $g_i, g_j$ are in involution if their corresponding derivations commute in $\dd$. In particular for $g_i$ and $f$, we have that $H$ and $H_i$ Poisson commute on the orbit if and only if $JA_i$ belongs to the centralizer of $JA$ in $\dd$, $z(JA)_{\dd}$. Since the complete integrability requires of n linearly independent functions, this can be done with a basis of an $n$-dimensional abelian subalgebra of $z(JA)_{\dd}$, finishing the proof.
\end{proof}

A linear map  $t$ is a derivation of $\hh_n$ acting trivially on the center $\zz(\hh_n)$ if and only if
$J t + t^* J=0$, if and only if
$t\in \mathfrak{sp}(n)$. The derivations of nilpotent Lie algebras of H-type were computed in (\cite{Sa}). 

In the case of the motion of n-uncoupled harmonic oscillators, we can see that the corresponding derivation is an element of a Cartan subalgebra of $\mathfrak{sp}(n)$.

\

\end{document}